\documentclass[fleqn,usenatbib]{mnras}
\usepackage[normalem]{ulem} 
\usepackage{newtxtext,newtxmath}
\usepackage{ulem}

\usepackage[T1]{fontenc}
\usepackage{ae,aecompl}
\usepackage{comment}

\DeclareRobustCommand{\VAN}[3]{#2}
\let\VANthebibliography\thebibliography
\def\thebibliography{\DeclareRobustCommand{\VAN}[3]{##3}\VANthebibliography}

\usepackage{stackengine,graphicx}	
\usepackage{amsmath}	

\usepackage{multirow}
\usepackage{multicol}
\usepackage{xcolor}
\newcommand{\sgra}{Sgr~A$^{\star}$}
\newcommand{\sgrb}{Sgr~B2}
\newcommand{\solM}{$M_{\odot}$}
\newcommand{\ka}{$K_{\alpha}$}
\newcommand{\xmm}{\textit{XMM--Newton}}
\newcommand{\intl}{\textit{INTEGRAL}}
\newcommand{\ib}{\textit{IBIS}}
\newcommand{\nus}{\textit{NuSTAR}}
\newcommand{\isg}{\textit{ISGRI}}
\newcommand{\igr}{IGR1747}
\newcommand{\oph}{Ophiuchus cluster}

\title[\sgrb\ hard X-ray emission with \intl\ after 2009]{\sgrb\ hard X-ray emission with \intl\ after 2009: still detectable?}

\author[Kuznetsova et al.]{Ekaterina Kuznetsova,$^{1}$\thanks{E-mail: eakuznetsova@cosmos.ru}
Roman Krivonos$,^{1}$ Alexander Lutovinov$,^{1,2}$ Ma{\"i}ca Clavel$^{3}$
\\
$^{1}$Space Research Institute of the Russian Academy of Sciences, Profsoyuznaya 84/32, 117997 Moscow, Russia\\
$^{2}$National Research University Higher School of Economics, Myasnitskaya str. 20, Moscow 101000, Russia\\
$^{3}$Univ. Grenoble Alpes, CNRS, IPAG, F38000 Grenoble, France\\
}

\date{Accepted XXX. Received YYY; in original form ZZZ}
\pubyear{2021}

\begin{document}
\label{firstpage}
\pagerange{\pageref{firstpage}--\pageref{lastpage}}
\maketitle


\begin{abstract}
Molecular cloud \sgrb\ is a natural Compton mirror in the Central Molecular Zone. An observed fading of the \sgrb\ X-ray emission in continuum and Fe~\ka\ 6.4~keV line indicates, as believed, a past X-ray flare activity of the supermassive black hole \sgra. The \sgrb\ was investigated by the \intl\ observatory at hard X-rays in 2003--2009, showing a clear decay of its hard X-ray emission. In this work, we present a long-term time evolution of the \sgrb\ hard X-ray continuum after 2009, associated with the hard X-ray source IGR~J17475--2822 as observed by \intl. The 30--80~keV sky maps, obtained in 2009--2019, demonstrate a significant excess spatially consistent with IGR~J17475--2822. The observed 2003--2019 light curve of IGR~J17475--2822 is characterized by a linear decrease by a factor of $\sim2$ until 2011, after which it reaches a constant level of $\sim1$~mCrab. The source spectrum above 17 keV  is consistent with a power-law model with $\Gamma=1.4$ and a high-energy cut-off at ${\sim}43$~keV. The \sgrb\ residual emission after $\sim$2011 shows a good correspondence  with models of the X-ray emission due to the irradiation of the molecular gas by hard X-rays and low-energy cosmic ray ions. We discuss the possible origin of the residual \sgrb\ emission after 2011 within these models, including theoretically predicted multiply-scattered emission.

\end{abstract}



\section{Introduction}

\sgra\ is a supermassive black hole (SMBH) with a mass of $\sim4.15\times10^6M_{\odot}$ distanced from us at 8.178~kpc \citep{Gravity2019}. At the moment, \sgra\ is in a quiescent state with an X-ray luminosity of $L_{\rm 2-10~keV}\sim2\times10^{33}$~ergs~s$^{-1}$ in the 2--10~keV energy band, which is ten orders of magnitude fainter than its Eddington luminosity predicted by the standard thin disk accretion onto a black hole (BH) \citep{Baganoff2003}. Such emission is also significantly lower than typical luminosities of active galactic nuclei (AGN) with comparable masses. It is natural to investigate whether \sgra\ was active with bright X-ray flares in the past.

\citet{Sunyaev1993} proposed the mechanism of reflection of strong X-ray flares from a low-ionized medium. This mechanism predicts hard X-ray emission with a continuum strongly absorbed at low energies and strong fluorescent Fe~\ka\ emission line at 6.4~keV with an equivalent width (EW) of about 1~keV from an X-ray reflection nebulae \citep{Sunyaev1993, Sunyaev1998}. Such emission is observed from the molecular clouds of the Central Molecular Zone \citep[CMZ, see][]{Morris1996} located in the Galactic Center (GC) region. The CMZ consists of $\sim10$\% of all molecular matter in the Galaxy and has a radius of about 200~pc. Its X-ray continuum and Fe~\ka\ emission line at 6.4~keV suggest the Compton mirror mechanism. A possible source of the strong hard X-ray emission which can produce the observed emission of the CMZ is a past flaring activity of \sgra. A large number of molecular clouds located in the GC region provides a unique possibility to investigate past activity of \sgra\ \citep{Sunyaev1998}. A variability of the molecular clouds' X-ray emission traces the propagation of the X-ray flare front from the regions closest to SMBH \sgra\ to the large distances from it \citep{Ponti2010}.

An alternative hypothesis of the GC molecular clouds X-ray emission is excitation of neutral matter by collisions with low-energy cosmic rays \citep[LECRs, see e.g.,][]{Dogiel2009,Dogiel2013,Dogiel2014}. LECRs can reproduce an X-ray continuum emission and Fe~\ka\ 6.4~keV emission line via bremsstrahlung and fluorescence mechanisms, respectively \citep[see e.g.,][]{Tatischeff2003, Tatischeff2012}. The observed fading of the molecular clouds' X-ray emission is in contradiction with the LECR hypothesis \citep[see e.g.,][]{Dogiel2014}. However, when the front of the X-ray flare finishes its propagation,  constant emission caused by LECRs may become visible. For example, \cite{Chernyshov2018} suggested a scenario of a combination of  reflected emission and emission excited by subrelativistic cosmic rays for a molecular cloud of the Arches cluster complex which emission was characterized by a fading non-thermal hard X-ray continuum and fluorescent Fe~\ka\ 6.4~keV line \citep{Krivonos2014,Krivonos2017,Clavel2014,Kuznetsova2019}.

\sgrb\ is the densest ($10^6$~cm$^{-3}$ in its 5~pc core) and most massive molecular cloud ($\sim10^6$~\solM) in the CMZ. Thanks to the {\it ASCA} observatory, the \sgrb\ emission in the fluorescent Fe~\ka\ line was detected and its high equivalent width (EW) was measured \citep{Koyama1996}. The \sgrb\ X-ray emission at energies above 20~keV was for the first time associated with the hard X-ray source IGR~J17475--2822 (hereafter \igr), detected with the \intl\ observatory by \cite{Revnivtsev2004}, who concluded that \sgrb\ could be irradiated by a hard X-ray flare from \sgra\ with a luminosity of $1.5\times 10^{39}$~erg~s$^{-1}$ in the 2--200~keV band and characterized by a spectral power-law photon index of $\Gamma\approx1.8$. Assuming the \sgrb\ projected distance from \sgra\ to be 100~pc, \citet{Revnivtsev2004} concluded that the \sgra\ flare occurred 300--400 years ago. Using \intl\ observations in 2003-2009, \cite{Terrier2010} obtained the light curve of \sgrb, showing a descending trend of its hard X-ray emission. The new parallax measurement of \sgrb\ obtained by \citet{Reid2009} suggested that \sgrb\ is 130~pc nearer than \sgra. \citet{Ponti2010}, considering the new \sgrb\ position, reported that the \sgra\ flare terminated 100~years ago. Using Monte Carlo simulations, \citet{Walls2016} considered two cases: uniform and Gaussian \sgrb\ density profiles. The former gave an estimate of the \sgrb\ position as being 50~pc closer to the Earth than \sgra, corresponding to an older flare than the estimation by \citet{Ponti2010}, while the latter suggested the \sgrb\ position at the projected distance of 100~pc that supports the 300--400-year-old flare. \nus\ observations of the \sgrb\ at energies up to 40~keV in 2013 allowed for the detection of prominent X-ray features and two compact cores in the central $90''$-region, which are surrounded by diffuse emission \citep{Zhang2015}. It is inconclusive whether the Fe~K$\alpha$ emission has reached a constant background level or is continuing to decrease. Moreover, the decreasing scenario is best explained with reflection, while a constant scenario is best described by cosmic rays.

Since 2013, there were no new investigations of the \sgrb\ emission at energies higher than 20 keV. However, the \intl\ observatory has continued its work, and a large amount of new data has been collected up to 2020. It is natural to expect that the \sgrb\ X-ray emission either continues to fade towards a non-detection level or reaches a constant level or rises due to another flare from \sgra. Indeed, a number of works suggest several \sgra\ flares in the past that are currently propagating in the CMZ \citep{Clavel2013,Chernyshov2018,Chuard2018,Terrier2018}.

In this work, we present the long-term evolution of the \sgrb\ hard X-ray emission obtained from the whole dataset of the \intl\ observations publicly available to date. The paper is structured as follows: Sect.~\ref{sec:obs} contains a brief overview of the \intl\ data processing and used observations. The time evolution of the \sgrb\ hard X-ray emission is presented in Sect.~\ref{sec:evol}. Sect.~\ref{sec:spec} describes the \sgrb\ spectral analysis. Discussion and summary are presented in Sect.~\ref{sec:dis} and \ref{sec:sum}, respectively.

\section{Observations and data analysis}
\label{sec:obs}

All publicly available \intl\ data from December 2002 to January 2020 were selected for this work. We used the coded-aperture telescope \ib\ \citep{2003A&A...411L.131U}, which operates in the soft gamma-ray energy band 20~keV -- 10~MeV on board the \intl\ observatory \citep{Winkler2003}. For our purposes, we utilized data from the low-energy detector layer \isg\ \citep{Lebrun2003} of the \ib\ telescope. Its spatial resolution of $12'$ (full width at half-maximum, FWHM) provides a possibility to detect \sgrb\ individually, which corresponds to a spatial scale of ${\sim}30$~pc at a distance of 8.5~kpc.

We used energy calibration as implemented in the \intl\ Science Data Center 'Off-line Scientific Analysis' (OSA) software version 10 for all data, but from the beginning of the 1626 revolution, OSA11 was applied to the remaining data (see details in  \url{https://www.isdc.unige.ch/integral/}). The \intl\ data were reprocessed with a proprietary analysis package developed at IKI\footnote{Space Research Institute of the Russian Academy of Sciences, Moscow, Russia} \citep[details available in][]{Krivonos2010,Krivonos2012,Churazov2014}. It was optimized for the \ib\ image deconvolution using the systematic noise suppression algorithm described in \citet{Krivonos2010}, which was proved to be effective in a number of works \citep[see e.g.,][]{Krivonos2012, Krivonos2017b}, that is crucial for the \sgrb\ image analysis in a crowded field of the GC region. For our analysis, we used individual sky images for each \intl\ observation with a typical exposure time of 2~ks (science window or ScW). To compensate the ongoing detector degradation, all ScW sky images were renormalized using the observed count rate of the Crab nebula, measured in the nearest observation, which provides smooth calibration of the ancillary response function over the whole observation time. The typical time between the \sgrb\ and Crab observations is about 30 days. As a result, final sky mosaics are constructed from the \ib\ ScW sky images in mCrab units.  Note that the described procedure automatically corrects for the intrinsic variability of the Crab Nebula flux (for details see Sect.~\ref{sec:syst}). For spectral analysis in this work, the diagonal energy redistribution matrix is designed to reproduce the Crab spectrum, which can be represented as $10.0\times E^{-2.1}$~keV~photons~cm$^{-2}$~s$^{-1}$~keV$^{-1}$ \citep[see, e.g.][]{Churazov2007, 2017ApJ...841...56M}. Note that the astrophysical (cosmic-ray X-ray and Galactic ridge backgrounds) and instrumental backgrounds of the \ib\ telescope are already subtracted as a result of the coded-mask image reconstruction algorithm, so the sky images have zero expectation value and unit variance.

We selected the 30--80~keV energy band for the analysis of the \sgrb\ long-term light curve since this band is characterized by an almost constant efficiency during more than a decade of \intl\ observations, i.e., it is not strongly affected by the ongoing detector degradation. Note that errors for all estimated parameters are given at the 90\% confidence interval.

\subsection{Systematic noise}
\label{sec:syst}

Before the detailed analysis of the \ib\ data, we checked them for presence of systematic noise in the flux determination. We used the Crab Nebula flux measured in each ScW in the following energy bands: 30--80~keV (for light curves, see the next paragraph) and 17--26, 26--38, 38--57, 57--86, 86--129~keV for the \sgrb\ spectra (see Sec.~\ref{sec:spec}). The relative systematic scatter of the measured fluxes was determined by dividing the standard deviation of the Crab flux by its average value during a given spacecraft revolution ($\sim$3 days). In all energy bands, systematic noise was found at a level of $\sim5-8\%$ over the considered time interval. The obtained level of systematic noise is comparable to the variability of the Crab Nebula flux observed at the 10\% level \citep[see, e.g.][]{Oh2018}. However, we found that in the 17--26~keV energy band, relative systematic noise was significantly increased from 7\% to $20-30\%$ after the $\sim1500$~revolution. The obtained systematic errors were taken into account in the following analysis.

Similar to \cite{Terrier2010}, we use the Ophiuchus cluster, located at $\sim9\degr$ from \sgrb, as a reference persistent X-ray source to verify our light curve extraction procedure. Fig.~\ref{fig:const} shows the $30-80$~keV multi-year light curve of the \oph\ approximated with a constant function at $F_{\rm Oph}=1.37\pm0.11$~mCrab\footnote{The flux unit of 1~mCrab is equivalent to the flux of $1.07\times10^{-11}$~erg~s$^{-1}$~cm$^{-2}$ in the 30--80~keV energy band for a source with a spectrum similar to that of the Crab Nebula.}. Despite the relative weakness of the \oph\ in the 30--80~keV band, we obtained  good fit statistics characterized by a reduced chi-squared $\chi^2_{\rm red}=1.4$ for 16 degrees of freedom (d.o.f.). We conclude that our \ib/\isg\ light curve extraction procedure in the 30--80~keV band is not significantly affected by systematic noise related to the multi-year \isg\ detector degradation.

\begin{figure}
\center
\includegraphics[width=1\columnwidth]{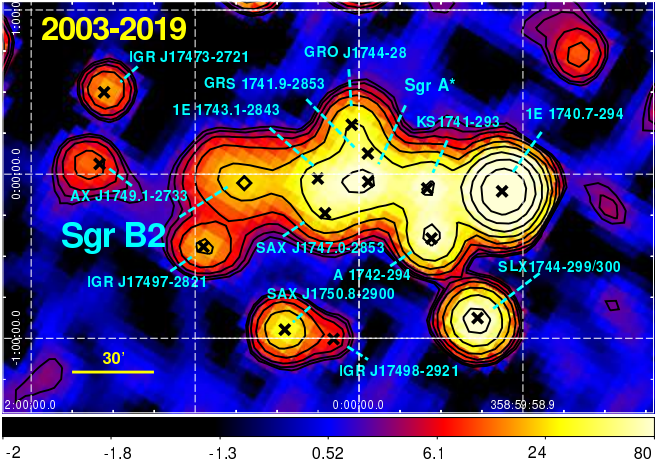}
\caption{2003--2019 (26--2180 \intl\ orbits) \ib\ image of the GC region in the 30--80~keV energy band shown in detection significance units. The positions of known bright X-ray sources are marked with crosses. Solid black contours denote significance levels from 2$\sigma$ to 800$\sigma$ in logarithmic scale. The \sgrb\ molecular cloud is shown by a diamond.}
\label{fig:map}
\end{figure}

\section{\ib/\isg\ sky imaging of \sgrb\ region}
\label{sec:map}

We first constructed the time-averaged 30--80~keV \ib/\isg\ map of the GC region using all available data from 2003 to 2020 (see Fig.~\ref{fig:map}). The \intl\ source \igr, spatially coincident with the position of \sgrb\ \citep{Revnivtsev2004b,2006ApJ...636..275B}, was significantly detected in the 30--80~keV band with the flux of $1.25\pm0.04$~mCrab at RA=$17^{\rm h}47^{\rm m}29\fs28$ and Dec.$=-28\degr21'57\farcs60$ (equinox J2000). The \igr\ centroid position is shifted by $\sim2\farcm1$ from the center of the peak of the \sgrb\ column density of the molecular gas localized between Sgr~B2(N) and Sgr~B2(M) cores at RA=$17^{\rm h}47^{\rm m}20\fs35$ and Dec.$=-28\degr22'43\farcs50$ \citep{Protheroe2008}. Note that the \intl/\ib\ point source localization accuracy depends on detection significance \citep{Krivonos2007}. Thus, the obtained offset is within the corresponding uncertainty of ${\sim}3'$ ($2\sigma$) for the source detection significance of $10-20\sigma$, which confirms the association of \igr\ with the \sgrb\ molecular cloud.

\begin{figure}
\center
\includegraphics[width=1\columnwidth]{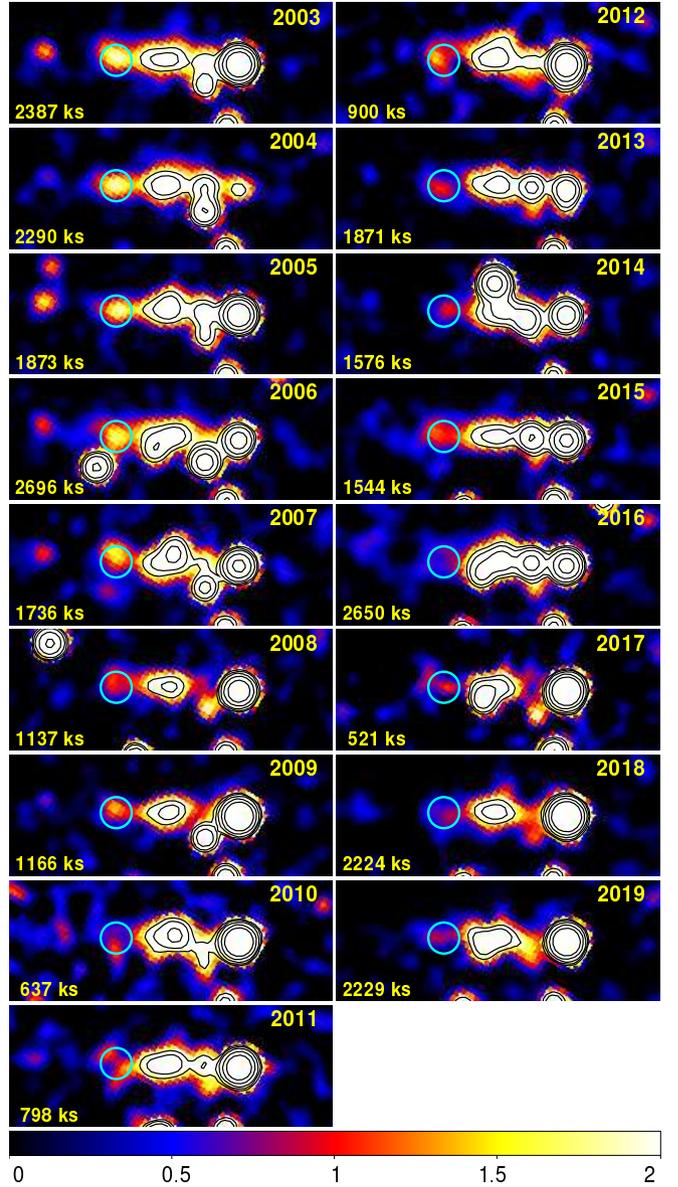}
\caption{\ib\ one-year averaged maps of central $4\degr$ in the 30--80~keV energy band shown in flux (mCrab units) from 2003 to 2019. Cyan circle with $R=6'$ reveals the position of the \sgrb\ source. The corresponding years and dead-time corrected exposures are indicated in panels. Contours denote isophotes of the flux levels of 2, 3, 5, 10, and 20~mCrab.}
\label{fig:years}
\end{figure}

The \ib\ one-year averaged maps of the GC~region shown in Fig.~\ref{fig:years} demonstrate a clear decrease in the \sgrb\ flux from 2003 to 2009, with the corresponding drop in significance from $13\sigma$ to $4\sigma$, which is consistent with \cite{Terrier2010} findings. On the sky maps after 2009, \sgrb\ appears as a weak source detected at a significance of $\sim2-5\sigma$. Note that apparent morphology changes can not be considered real because the \intl\ localization accuracy of a weak source is $4.2'$ at the $2\sigma$ confidence interval \citep{Krivonos2007}. Also from Fig.~\ref{fig:years}, it is seen that some GC sources have strong flux variations with a timescale of less than 1~year. Contours denote a surface brightness above 2~mCrab in order to better represent of the nearby X-ray sources with the flux higher than that of \sgrb. Fig.~\ref{fig:years} shows that the hard X-ray emission at the position of \sgrb\ continues to be visible after 2010.

\section{Long-term light curve of \sgrb}
\label{sec:evol}

\begin{figure}
\center
\includegraphics[width=1\columnwidth]{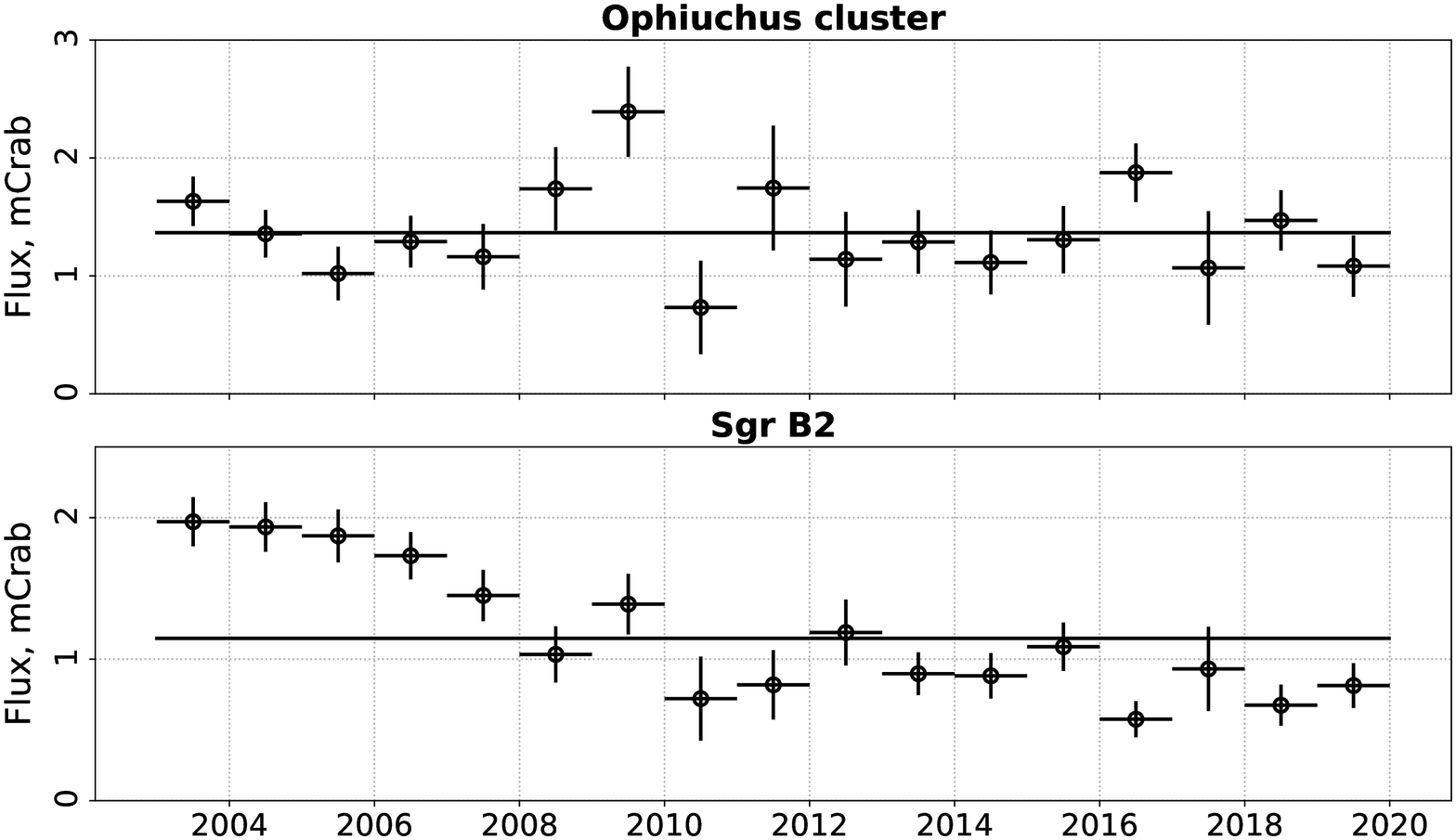}
\caption{The light curves of the \oph\ (upper plot) and \sgrb\ (bottom plot), obtained by \ib\ in 2003--2019. Solid lines represent the best-fit constant functions for the \oph\ and \sgrb\ light curves, shown in the upper and bottom plots, respectively.}
\label{fig:const}
\end{figure}

The obtained 2003--2020 \ib\ light curve of \sgrb\ is shown in Fig.~\ref{fig:const}. The fitting procedure applied to the light curve with a simple constant function does not describe the data well enough, as reflected in the worst fit statistics $\chi^2_{\rm red}/d.o.f.=7.4/16$ (see Table~\ref{tab:lin}), which is mainly caused by a clear decay of the \sgrb\ hard X-ray emission before 2009, as reported by \cite{Terrier2010}. Therefore, the Sgr B2 flux is inconsistent with the constant function and demonstrates a few-year time variability. Note that a possible fast Sgr B2 variability, if present, is diluted within the one-year time bins, and on smaller time bins, the data quality does not allow us to draw any firm conclusions.

\begin{figure}
\center
\includegraphics[width=1\columnwidth]{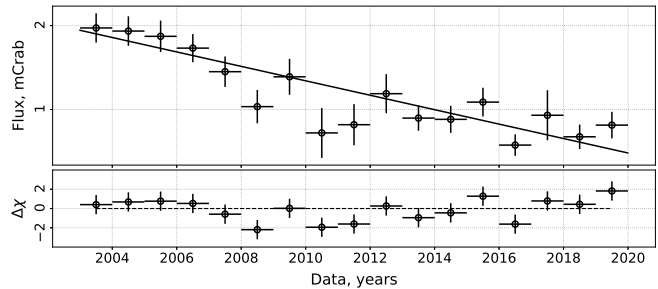}
\caption{The light curve of \sgrb\ in the 30--80~keV energy band approximated with the linear function (upper plot) and residuals (bottom plot).}
\label{fig:lin}
\end{figure}

To describe the decreasing trend of the \sgrb\ emission with an additional background component, we applied a linear function of the form $A*T+B$, where A and B are coefficients in units of mCrab/yrs and mCrab, correspondingly, and T is a time in years since 2003. This linear model provides a better fitting of the \sgrb\ light curve, as shown in Fig.~\ref{fig:lin} and listed in Table~\ref{tab:lin}. To compare our results with that obtained by \cite{Terrier2010}, we chose the time parameter $\tau_{1/2}$, a time of flux decrease by a factor of two from the flux initial value. Our estimation of $\tau_{1/2}=12\pm2$~yrs obtained within the time interval of 2003--2020 is somewhat larger than $\tau_{1/2}=8.2\pm1.7$~yrs determined by \cite{Terrier2010} in 2003--2009. The difference is probably caused by the longer fitting interval considered in the current work.

\begin{figure}
\center
\includegraphics[width=1\columnwidth]{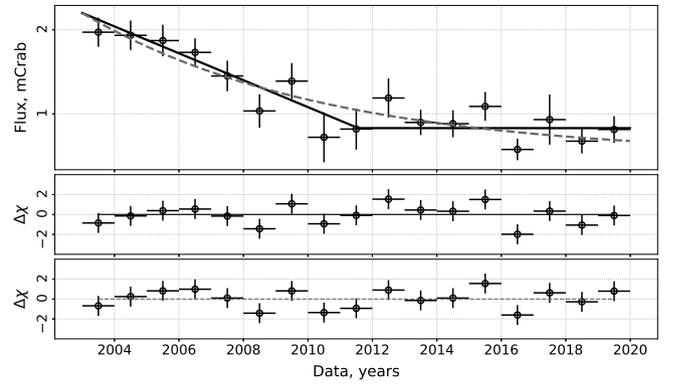}
\caption{The 30--80~keV light curve of \sgrb\ approximated with the linear piece-wise and exponential functions, as shown in the upper plot, respectively, with solid and dashed lines, and the corresponding residuals are demonstrated in the middle and bottom panels.}
\label{fig:pw}
\end{figure}

\begin{table}
\noindent
\centering
\caption{Best-fit parameters for the fitting procedure of the \sgrb\ light curve with the constant, linear, piece-wise, and exponential functions.}
\label{tab:lin}
\vspace{1mm}
\begin{tabular}{c|c|c|c|c}
\hline \hline
Parameter                 & Constant  & Linear          & Piece-wise     & Exponential         \\
\hline
A, mCrab/yrs              & ---       & $-0.08\pm0.01$  & $-0.16\pm0.06$ &  ---                \\
B, mCrab                  & ---       & $1.9\pm0.14$    & $2.2\pm0.2$    &  ---                \\
$T_{\rm break}$, yrs      & ---       & ---             & $2011\pm3$     &  ---                \\
$\tau_{1/2}$, yrs         & ---       & $12\pm2$        & $6\pm2$        &  ---                \\
C, mCrab                  & $\sim1.1$ & ---             & $0.83\pm0.1$   & $0.5^{+0.3}_{-0.6}$ \\
$F_0$, mCrab              & ---       & ---             & ---            & $1.7^{+0.5}_{-0.3}$ \\
$\tau$, yrs               & ---       & ---             & ---            & $7.5^{+7.1}_{-3.5}$ \\
$\chi^2_{\rm red}/d.o.f.$ & 7.4/16    & 1.48/15         & 1.09/14        & 1.04/14             \\
\hline
\end{tabular}\\
\vspace{3mm}
\end{table}

Then we suggested that there is a constant level in the \sgrb\ decreasing trend described by a linear piece-wise function in the following form:
\begin{equation}
F(T) =  \begin{cases}
				A*T+B~for~T \leq T_{\rm break} \\
                C~for~T > T_{\rm break} \\
        \end{cases}
            ,
\end{equation}
where $T_{\rm break}$ is a time when the \sgrb\ emission changes from a linear trend to a constant $C$ level. The approximation of the light curve is shown in Fig.~\ref{fig:pw}, with the best-fitting model parameters listed in Table~\ref{tab:lin}. The piece-wise function describes the \sgrb\ light curve at better fit statistics compared to the constant and linear models. The characteristic time $\tau_{1/2}=6\pm2$~yrs is in agreement with the \citet{Terrier2010} estimation within the uncertainties. Note that the constant level $C=0.8\pm0.1$~mCrab measured after $T_{\rm break}=2011\pm3$ is not consistent with zero flux background level, as it is expected for the coded-mask sky reconstruction method \citep[see e.g.,][and references therein]{Krivonos2010}.

We divided the whole dataset from 2003 to 2019 into two time intervals T1 and T2, separated by the year 2011 (orbit 1055), where a linear decay is replaced by a constant according to the linear piece-wise fit (Sect.~\ref{sec:evol}). Fig.~\ref{fig:map_2} demonstrates the T1 and T2 ficance maps in the 30--80~keV energy band. The \sgrb\ 30--80~keV flux was measured at the level of 1.6~mCrab ($23.5\sigma$) and 0.8~mCrab ($10.7\sigma$), respectively, for the T1 and T2 datasets.

The \sgrb\ centroid positions at  RA=$17^{\rm h}47^{\rm m}28\fs32$, Dec.$=-28\degr21'10\farcs80$ and RA=$17^{\rm h}47^{\rm m}25\fs75$, Dec.$=-28\degr25'08\farcs18$ measured during the T1 and T2 time intervals are shifted, respectively, by $2.3'$ and $2.7'$ from the position of the maximum column density of the cloud \citep{Protheroe2008}. The observed offsets are well consistent with the $3'$ localization uncertainty ($2\sigma$) for weak sources detected by \intl/\ib\ \citep{Krivonos2007}.

Finally, we approximated the light curve of \sgrb\ by an exponential decay with a constant term $C$: $F_0 \times \exp^{-\lambda T}+C$, where $F_0$ is an initial flux in mCrab units, $\tau$ is a lifetime expressed in the units of yrs. The best-fitting model describes the light curve at acceptable fit statistics $\chi^2_{\rm red}/d.o.f.=1.04/14$, as listed in Table~\ref{tab:lin} and shown in Fig.~\ref{fig:pw}. The lifetime was estimated at $\tau=7.5^{+7.1}_{-3.4}$~yrs, which is close to $\tau\sim11$~yrs provided by \cite{Zhang2015} for the central $90''$ of \sgrb.

\section{Spectral analysis}
\label{sec:spec}

\begin{figure}
\center
\includegraphics[width=1\columnwidth]{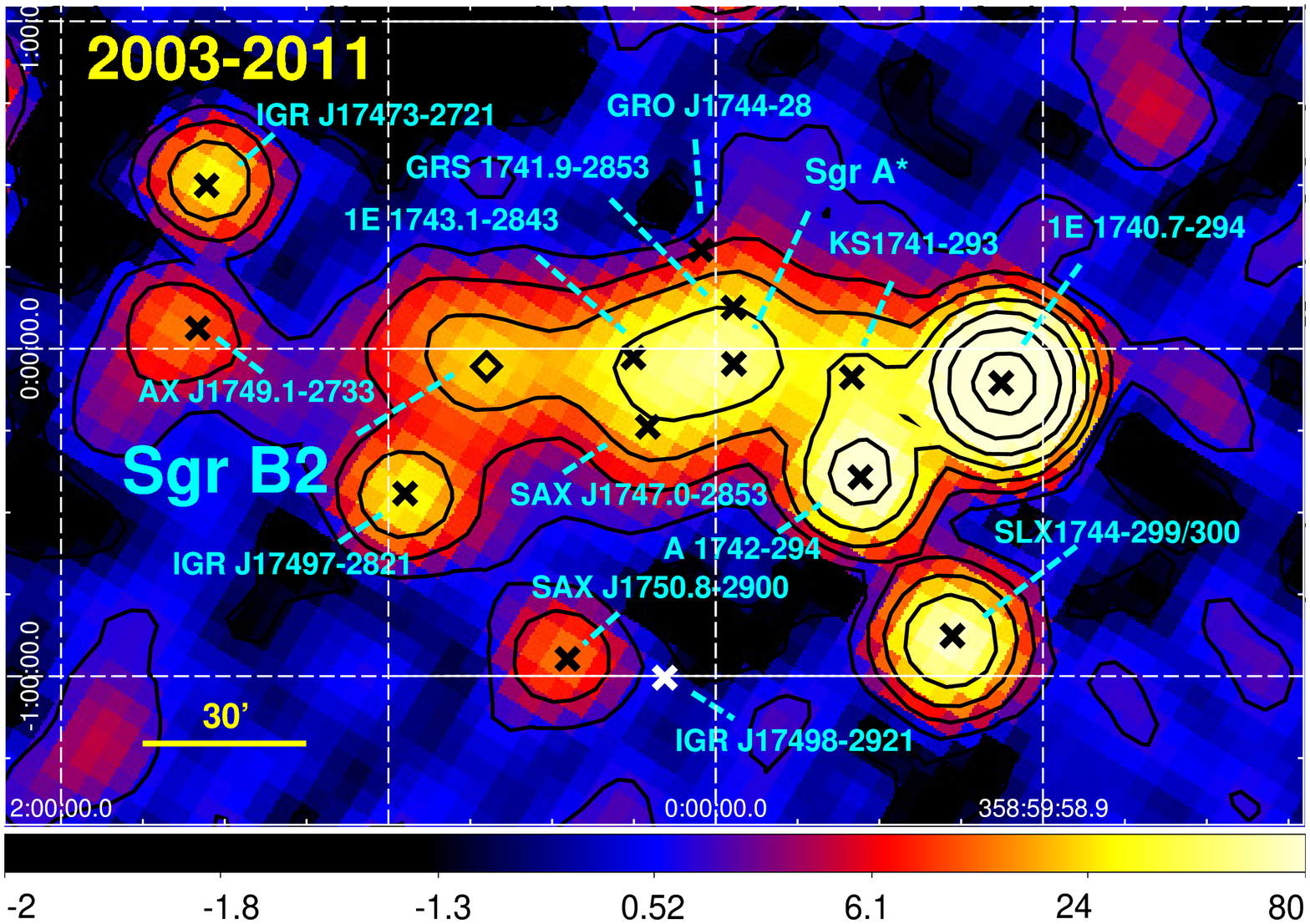}
\includegraphics[width=1\columnwidth]{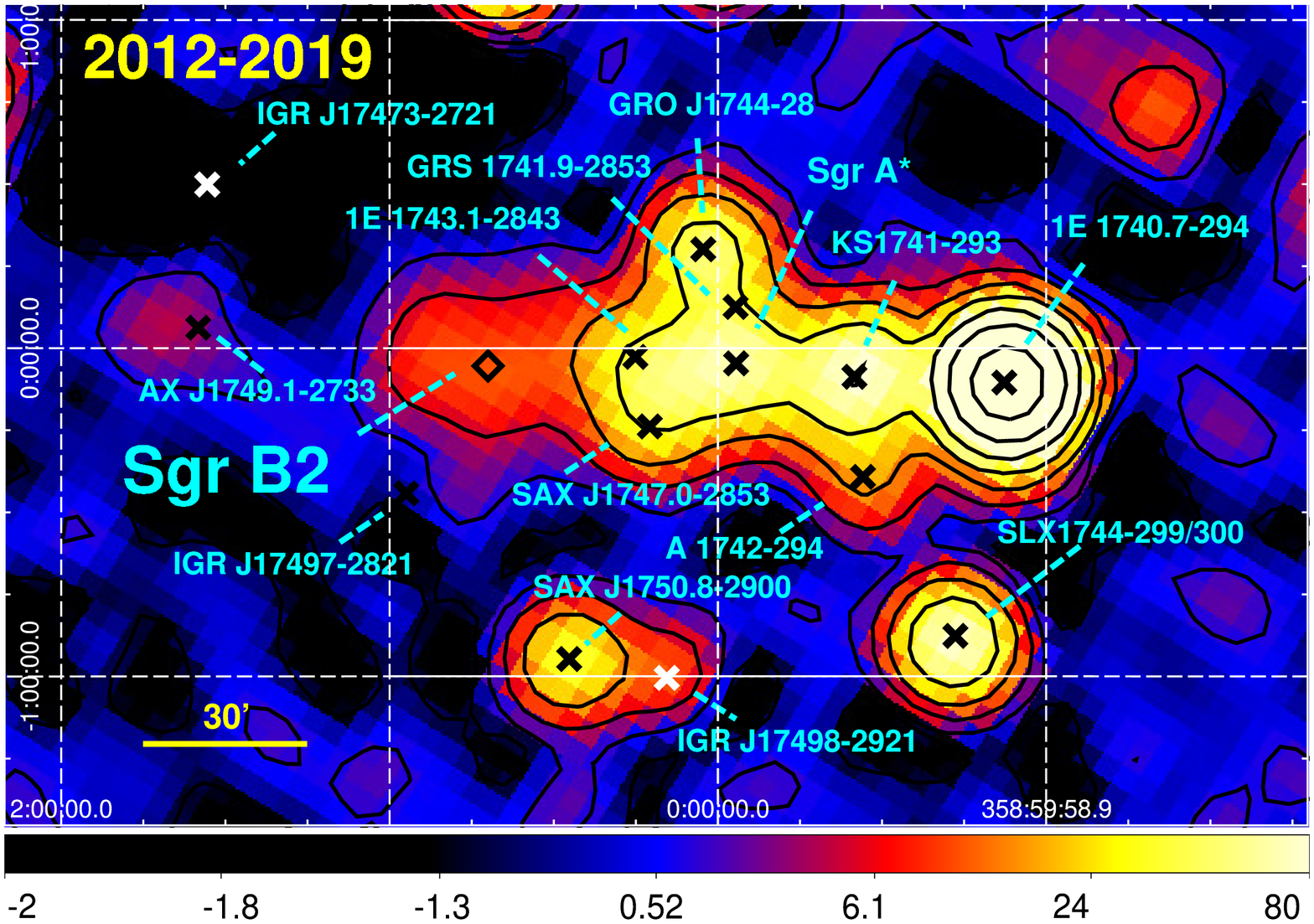}
\caption{\ib\ images of the GC region in the 30--80~keV energy band in detection significance units for the T1 and T2 time intervals are in the top and bottom panels, respectively. Contours and regions are the same as in Fig.~\ref{fig:map}.}
\label{fig:map_2}
\end{figure}

In this section, we analyse spectral information of \sgrb\ in the wide energy range 17--129~keV obtained for the T1 and T2 time intervals (Sect.~\ref{sec:evol}). We extracted spectra from the sky images in different energy bands in the position of \sgrb, as shown in Fig.~\ref{fig:map_2}. Note that the region of spectral extraction corresponds to the angular resolution of \ib/\isg\ of $12'$. We excluded the first energy bin from the T2 spectrum due to the ongoing \isg\ detector degradation and loss of sensitivity at low energies \citep{Caballero2013}. Note that we also added a 5\% systematic error in the spectral fitting procedures.

\begin{figure}
\center
\includegraphics[width=1\columnwidth]{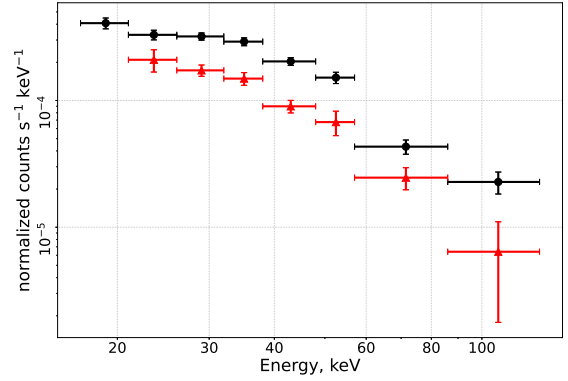}
\caption{The \sgrb\ region spectra obtained in the 17--129~keV (black circles) and 21--129~keV (red triangles) energy band by \ib\ during T1 (circles) and T2 (triangles).}
\label{fig:2spec}
\end{figure}

\begin{figure}
\center
\includegraphics[width=1\columnwidth]{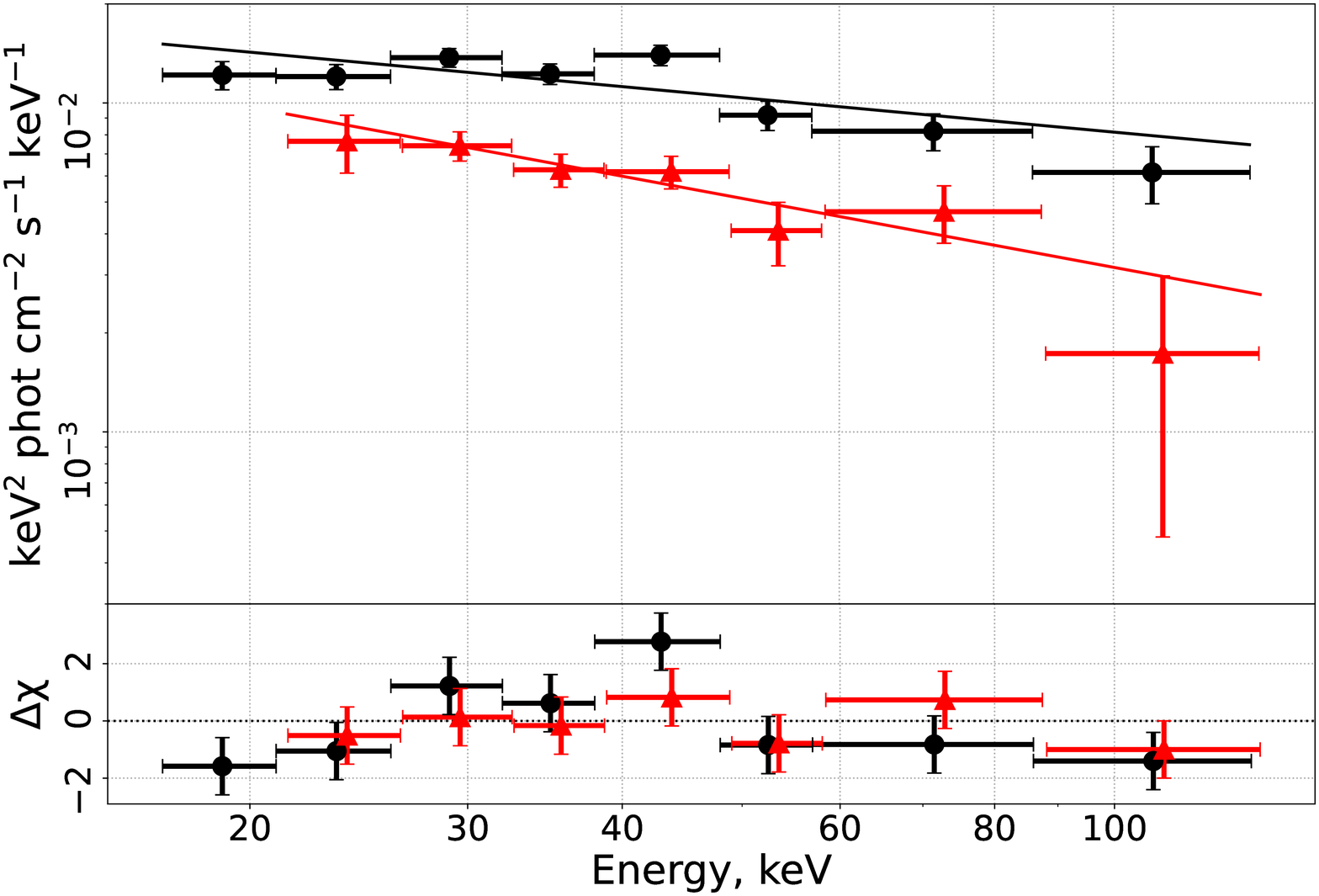}
\includegraphics[width=1\columnwidth]{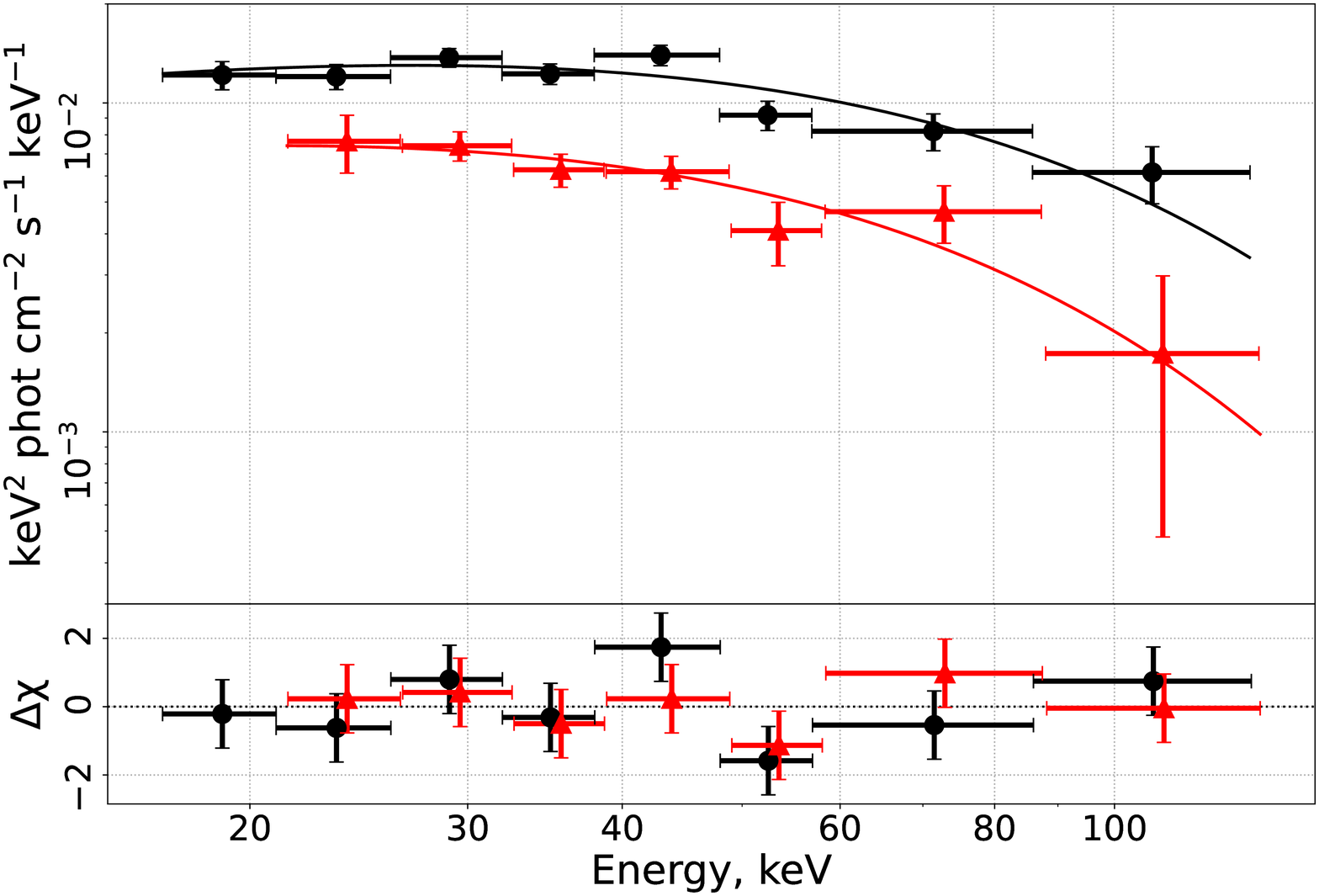}
\caption{The \sgrb\ region spectra obtained in the 17--129 and 21--129~keV energy bands by \ib\ during T1 (circles) and T2 (triangles), respectively. Black and red solid lines represent the best-fit simple power-law (top panel) and cutoff power-law (bottom panel) models for the \sgrb\ spectra obtained before and after $T_{\rm break}$, respectively. For convenience, in Fig.~\ref{fig:spec}, we do not present the complex models $CREFL16$ and $LECRp$.}
\label{fig:spec}
\end{figure}

First, we tried to fit the obtained \sgrb\ region spectra (see Fig.~\ref{fig:2spec}) with a power-law model. For spectral fitting, we used the {\sc xspec} package version 12.11.0 \citep{Arnaud1996}, which is part of the {\sc heasoft v6.27} software. The results are listed in Table~\ref{tab:spec}, where $\Gamma$ is a photon index of the power-law model. For this and all following models, we calculated the 25--50~keV flux $F_{\rm 25-50~keV}$ using the multiplicative model component $cflux$ in {\sc xspec}. The power-law model is in agreement only with the T2 spectrum (see Table~\ref{tab:spec}), and its $\Gamma$ is significantly greater than $\Gamma=2.07\pm0.05$ obtained for the hard X-ray part of the broad-band analysis in \cite{Terrier2010}. The upper plot in Fig.~\ref{fig:spec} shows the \sgrb\ region spectra fitted with the power-law model. Note that residuals of the T1 spectrum show evidence for a high-energy cutoff at ${\sim}40$~keV.

Then we fitted both spectra using a power-law model with a high-energy cutoff. The best-fitting model parameters are listed in Table~\ref{tab:spec}. The \sgrb\ region T1 spectrum is in good agreement with this model with a cutoff energy of $E_{\rm cut}=44^{+61}_{-18}$~keV (see bottom panel in Fig.~\ref{fig:spec}). Due to the low quality of the T2 data, the photon index is not constrained by the fit, so we fixed it at the previously obtained value $\Gamma=1.4$. We conclude that for the T1 spectrum, the cutoff power-law provides a better approximation than the simple power-law (see Table~\ref{tab:spec}).

Assuming that the spectral shape has not changed from the T1 time interval to T2, we jointly fitted both spectra with the cutoff power-law model (see column Joint in Table~\ref{tab:spec}). Here $C_{\rm cross}$ is a normalization constant of the T2 spectrum with respect to T1. The measured $C_{\rm cross}$ value of $0.51\pm0.06$ shows that the T1 flux has dropped by a factor of $\sim2$. The best-fit result has good statistics that indicate a possible similarity between the T1 and T2 spectral shapes.

\begin{table}
\noindent
\centering
\caption{Best-fit parameters of different models after the fitting procedure of the \sgrb\ region spectra obtained in the T1 and T2 time intervals.}
\label{tab:spec}
\vspace{1mm}
\begin{tabular}{c|c|c|c}
\hline \hline
Parameters                             & T1                     & T2                  & Joint               \\
\hline
\multicolumn{4}{c}{Power-law}                                                                               \\
\hline
$\Gamma$                               & $\sim$2.3              & $2.7\pm0.3$         & ---                 \\
$F^{\rm pow}_{\rm 25-50~keV}$,         & $\sim$13               & $7.2\pm0.7$         & ---                 \\
$10^{-12}~\rm erg~cm^{-2}~s^{-1}$      &                        &                     &                     \\
$C_{cross}$                            &                        &                     & ---                 \\
$\chi^2_{\rm red}/d.o.f.$              & 2.75/6                 & 0.63/5              & ---                 \\
\hline
\multicolumn{4}{c}{Cutoff power-law}                                                                        \\
\hline
$\Gamma$                               & $1.4\pm0.6$            & $1.4$ (fixed)       & $1.4^{+0.5}_{-0.6}$ \\
$E_{\rm cut}$, keV                     & $44^{+61}_{-18}$       & $35^{+12}_{-8}$     & $43^{+48}_{-17}$    \\
$F^{\rm pow}_{\rm 25-50~keV}$,         & $14\pm1$               & $7.3^{+0.7}_{-0.8}$ & $13.9\pm0.9$        \\
$10^{-12}~\rm erg~cm^{-2}~s^{-1}$      &                        &                     &                     \\
$C_{cross}$                            &                        &                     & $0.51\pm0.06$       \\
$\chi^2_{\rm red}/d.o.f.$              & 1.51/5                 & 0.55/5              & 1.08/11             \\
\hline
\multicolumn{4}{c}{$CREFL16$}                                                                               \\
\hline
$\Gamma$                               & $2.30\pm0.13$          & 2.3 (fixed)         & $2.35\pm0.12$       \\
$F^{\rm pow}_{\rm 25-50~keV}$,         & $13.6\pm0.8$           & $7.0\pm0.7$         & $13.6\pm0.8$        \\
$10^{-12}~\rm erg~cm^{-2}~s^{-1}$      &                        &                     &                     \\
$C_{cross}$                            &                        &                     & $0.52\pm0.06$       \\
$\chi^2_{\rm red}/d.o.f.$              & 1.53/6                 & 0.85/6              & 1.14/12             \\
\hline
\multicolumn{4}{c}{$LECRp$}                                                                                 \\
\hline
$\Gamma$                               & ---                    & $2.7\pm0.5$         & ---                 \\
$N$,$10^{-6}~\rm erg~cm^{-2}~s^{-1}$   & ---                    & $2.7^{+3.5}_{-1.3}$ & ---                 \\
$\chi^2_{\rm red}/d.o.f.$              & ---                    & 0.87/5              & ---                 \\
\hline
\end{tabular}\\
\vspace{3mm}
\end{table}

Then we used the physically motivated $CREFL16$ {\sc XSPEC} table model \citep[see description in][]{Churazov2017}, which describes a spectrum of a uniform gas cloud illuminated by an external source of a parallel X-ray emission beam. This model represents the reflected emission from the GC molecular clouds. The reflected emission depends on five parameters: a radial Thomson optical depth of the cloud $\tau_T$, a slope of the primary power-law spectrum $\Gamma$, an abundance of heavy elements relative to \citet{Feldman1992} $Z/Z_{\odot}$, a cosine of the viewing angle $\mu=cos\Theta$, and a normalization. We fixed $\tau_T$ = 0.4, $Z/Z_{\odot}$ = 1.9 in accordance with \cite{Revnivtsev2004} and considered the case in which \sgrb\ is in the same plane with \sgra, i.e., $\mu=cos~90\degr=0$. (see Table~\ref{tab:spec}). In this configuration, the $CREFL16$ model well describes the T1 spectrum with $\Gamma=2.30\pm0.13$. The T2 spectrum does not allow us to estimate $\Gamma$, so we fixed it at the T1 value assuming the same flare for both time intervals and obtained good fit statistics (see Table~\ref{tab:spec}).

Alternatively to the reflection mechanism, the \sgrb\ X-ray emission may be caused by interaction of a molecular cloud material with cosmic ray particles. \citet{Tatischeff2012} studied a non-thermal emission of a neutral ambient medium caused by LECR electrons and protons (hereafter LECRe and LECRp, respectively) and developed corresponding spectral models. We tested only the LECRp model due to the unphysical parameters of the LECRe model determined for the \sgrb\ emission by \citet{Zhang2015}. The LECRp model depends on five parameters: a power-law slope of an accelerated cosmic ray (CR) spectrum $s$, a minimum energy of the CRs $E_{\rm min}$, a metallicity of the ambient medium $Z/Z_{\odot}$, a path length of the CRs in the region $\Lambda$, a normalization of the model $N_{\rm LECRp}$. We fixed $Z/Z_{\odot}=2.5$, $\Lambda=5\times10^{24}$~H-atoms~cm$^{-2}$, $E_{\rm min}=10$~MeV in accordance with \citet{Zhang2015}, while $N_{\rm LECRp}$ and $s$ were free parameters. Due to the time variability in the \sgrb\ light curve at a few-year scale obtained for the T1 time interval (see Sect.~\ref{sec:evol}), which contradicts the CR scenario \citep[see e.g.,][]{Dogiel2014,Tatischeff2012}, we applied this model only at the T2 spectrum. The obtained results are in Table~\ref{tab:spec}.

\section{Discussion}
\label{sec:dis}

The decreasing behavior of the \sgrb\ X-ray emission was observed with many X-ray telescopes \citep{Inui2009, Terrier2010, Nobukawa2011, Zhang2015, Terrier2018}. The most suitable hypothesis, explaining the time variability of the \sgrb\ hard X-ray emission, is the reflection of the SMBH \sgra\ X-ray flare. However, the question of how long the \sgrb\ emission will keep its decreasing trend and which mechanism will dominate the observed X-ray emission after the \sgra\ light front leaves the cloud remains open.

Previous observations of the \sgrb\ region by X-ray observatories \nus\ and \xmm\ showed that the Fe~\ka\ flux of the central $90''$ region in 2013 is consistent with the flux measured in 2012, but at the same time the 2013 flux is also consistent with the decreasing trend \citep{Zhang2015}. These authors concluded that if the \sgrb\ emission has reached its background level, LECRp may be the main contributor. Although if the \sgrb\ flux continues to decrease, the reflection scenario better describes the observed emission behaviour. Also, \cite{Terrier2018} showed that in 2012 an extended X-ray emission from the \sgrb\ region was still detected with \xmm.

A similar case was observed in a molecular cloud near the Arches stellar cluster in the GC. Both fluxes of the non-thermal continuum and the Fe~\ka\ emission line demonstrate a decreasing trend with a similar timescale \citep{Clavel2014,Krivonos2017}, however, also showing evidence for the constant emission level \citep{Kuznetsova2019}. Also, \citet{Chernyshov2018} considered two scenarios for the Arches cluster molecular cloud emission: two X-ray flares $\sim$100~years and $\sim$200~years ago, which were reflected from two different clouds located on the line-of-sight, and a combination of reflected emission and emission caused by the bombardment of the molecular cloud matter by cosmic rays. The first scenario strongly depends on the reflection geometry of two clouds and the iron abundances, while the second one is restricted by the photon index of the X-ray emission and needs the presence of a local cosmic ray particle accelerator. Also, note that the variability of the X-ray emission at different time scales of several years observed from other molecular clouds in the GC region \citep[see e.g.,][]{Ponti2010, Ponti2013, Clavel2013, Ryu2013, Churazov2017, Terrier2018} may be related to different light-crossing times and/or intrinsic flare durations.

The \intl\ observatory allows us to collect information about the \sgrb\ region during the 17~years since 2003. Spatially consistent with \sgrb, the hard X-ray source \igr\ was detected during the considered time interval. The obtained 30--80~keV light curve of \igr\ shows a linear decay, which changed to a constant level in 2011, that separated the time interval 2003--2020 into the T1 and T2 epochs. This result is in agreement with the assumption that the \sgrb\ flux in 2013 remains at the 2012 level \citep{Zhang2015}. Before 2011, the decay is characterized with $\tau_{1/2}=6\pm2$~yrs that is consistent with the 2003--2009 \intl\ result $\tau_{1/2}=8.2\pm1.7$~yrs \citep{Terrier2010}. Thus the \intl\ light curve supports the reflection scenario for the time interval before $\sim2011$. Assuming the dominating single scattering scenario over the whole studied period 2003--2020, characterized by a single linear trend, $\tau$ is estimated to be $12\pm2$~yrs. This value is slightly larger than that obtained by \citet{Terrier2010}. It is worth noting that the decrease in the \sgrb\ hard X-ray continuum observed by \intl\ is in agreement with the overall drop of the 6.4~keV line flux measured by \xmm\ in the same sky region \citep{2021arXiv210813399R}. This indicates that the Fe K$_{\alpha}$ line flux and the hard X-ray continuum are linked to each other, confirming the reflection scenario.

The spectral analysis also points to the reflection origin because the 17--129~keV spectrum agrees with the $CREFL16$ model. The \intl\ T1 estimation of the initial flare slope $\Gamma=2.30\pm0.13$ (at the 90\% confidence interval) is in agreement with $\Gamma\sim2$ obtained for the \sgrb\ $90''$ central region by \nus\ \citep{Ponti2010, Mori2015, Zhang2015}. Also, \citet{Revnivtsev2004} reported about the power-law with $\Gamma=1.8\pm0.2$ (at the $1\sigma$ confidence interval) for the \sgrb\ spectrum using the \intl\ data collected in the 2003--2004 time interval. Furthermore, the similar slope $\Gamma\sim2$ was estimated for other GC molecular clouds by the reflected emission \citep{Mori2015, Krivonos2017} and for the observed \sgra\ X-ray flares \citep{Baganoff2001, Porquet2003,  Porquet2008, Nowak2012, Degenaar2013, Neilsen2013, Barriere2014, Zhang2017}.

The LECR hypothesis can be rejected for this emission due to the \sgrb\ multi-year time variability. Point sources may be additional contributors to the T1 \igr\ flux, but not the main ones because we did not detect any fast variability and the summed flux from the point sources did not describe all flux from the \sgrb\ region (see Sect.~\ref{sec:point}). Therefore, the reflection scenario is the most suitable for the \sgrb\ emission before 2011.

The main question is the origin of the observed constant emission after 2011 when the \sgrb\ flux dropped by a factor of ${\sim}2$ and remained at this level until 2019. We did not detect any significant difference between the T1 (before 2011) and T2 (after 2011) spectral shapes of the \sgrb\ emission. Is the \igr\ hard X-ray emission after 2011 still related to \sgrb\ molecular cloud or not?

\subsection{Reflection scenario after 2011}

Assuming that the light front from the \sgra\ flare has mostly left the cloud, another possible explanation of the observed remaining \sgrb\ emission is caused by a long-lived doubly-scattered albedo \citep{Sunyaev1998, Odaka2011, Molaro2016}, which probably became visible after 2011. The doubly-scattered albedo is expected to dominate single scatterings in hard X-rays due to a low photo-absorption and an additional enhancement by the down-scattering of high energy photons due to the Compton effect \citep{Sazonov2020, 2020MNRAS.495.1414K}.

\subsection{Cosmic ray scenario}

The constant T2 flux may be a sign of an interaction of the molecular cloud neutral matter with CRs. \cite{Zhang2015} suggested that in the case of the Fe~\ka\ emission of the $90''$ \sgrb\ region having reached its constant level in 2013, LECRp may be a major contributor. \cite{Zhang2015} rejected the LECRe hypothesis due to the unphysically large metallicity $Z/Z_{\odot}=4.0^{+2.0}_{-0.6}$ and very low electron energy. The high value of the photon index $\Gamma=2.7\pm0.3$ determined for the \intl\ T2 data obtained in the 21--129~keV energy band can be easily explained by the LECRp hypothesis, rather than the LECRe one \citep[see~Fig. 7 and 9 for the LECRp and Fig.~2 and 4 for the LECRe in ][]{Tatischeff2012}. A good agreement between the \intl\ T2 spectrum and the LECRp model points to the possible LECRp nature of the T2 \sgrb\ region spectrum.

Assuming that \sgrb\ is in the plane of \sgra, i.e., at the $D=8.178$~kpc distance \citep{Gravity2019}, we estimated the power injected by the LECR 10~MeV -- 1~GeV protons in the \sgrb\ region $dW/dt=4\pi D^2N_{\rm LECR}=2.1^{+2.6}_{-1.0}\times10^{40}$~erg~s$^{-1}$. This value is significantly greater ($\sim10$ times) than the \nus\ estimation \citep{Zhang2015}, probably due to the larger \intl\ spectrum extraction region. Then, taking into account the uncertainties of $E_{\rm min}$ \citep[see Sect. 5.2 in][]{Tatischeff2012}, we obtained $dW/dt=(4-90)\times10^{39}$~erg~s$^{-1}$. Note that the estimated slope of the CR source spectrum $s=2.7\pm0.5$ is in contradiction with the slope $1.5<s<2$ predicted by the diffusive shock acceleration (DSA) theory, which was considered in \cite{Tatischeff2012}. Therefore, we do not take into account the contribution from the suprathermal protons and possible heavier nuclei, and the estimation $dW/dt=(4-100)\times10^{39}$~erg~s$^{-1}$ is only a lower limit of the total CR power.

To investigate the ionization rate, we first determined the CR power deposited into the \sgrb\ cloud. The previously obtained LECRp power is $dW/dt\sim2.1\times10^{40}$~erg~s$^{-1}$. However the power deposited into the cloud is lower than the injected $dW/dt$ due to two reasons: (1) the non-penetration into the cloud of the CRs with $E<E_{min}$ and (2) the escape from the cloud of the CRs with the highest energies \citep{Tatischeff2012}. Note that we considered only protons and did not do any corrections of the $dW/dt$ related to the first reason because we are already taking into account only the CR particles with $E>E_{\rm min}$. For   $\Lambda=5\times10^{24}$~H-atoms~cm$^{-2}$, the protons with the energy $E>180$~MeV are not stopped in the cloud \citep{Tatischeff2012}. Integrating the CR power taking into account only the CRs with $E>10$~MeV/nucleon and $E<180$~MeV/nucleon, we obtained that the power deposited by LECRs into the cloud is $\dot{W_d}\sim1.9\times10^{40}$~erg~s$^{-1}$ (90\% of the $dW/dt$). Using the total mass of the \sgrb\ molecular cloud $M=6\times10^6M_{\odot}$ \citep{Lis1990}, we estimated the CR ionization rate at $\zeta_{\rm H}\sim5.6\times10^{-14}$~H$^{-1}$~s$^{-1}$ \citep[see Equation~11 in][]{Tatischeff2012}. Such a CR ionization rate is too high compared with the GC CR ionization rate $\zeta_{\rm H}\sim(1-3)\times10^{-15}$~H$^{-1}$~s$^{-1}$ \citep{Goto2011} and slightly higher than the \cite{Zhang2015} estimation $\zeta_{\rm H}\sim(6-10)\times10^{-15}$~H$^{-1}$~s$^{-1}$. Due to the high predicted CR ionization rate, we consider the LECRp scenario to be unfavorable.

\citet{Dogiel2011} predicted that in the case of the subrelativistic protons produced by accretion processes onto the SMBH \sgra, two X-ray emission components from the molecular cloud should be observed: a time variable reflected emission and a quasi-stationary emission caused by these protons. The authors predicted the time variations of the 6.4~keV Fe~\ka\ equivalent width when the \sgra\ flare front passes through the \sgrb\ cloud and the reflected flux completely drops. Future observations of \sgrb\ 6.4~keV line emission will shed light on the nature of its emission.

\subsection{Unresolved point sources scenario}
\label{sec:point}

\begin{figure}
\center
\includegraphics[width=0.8\columnwidth]{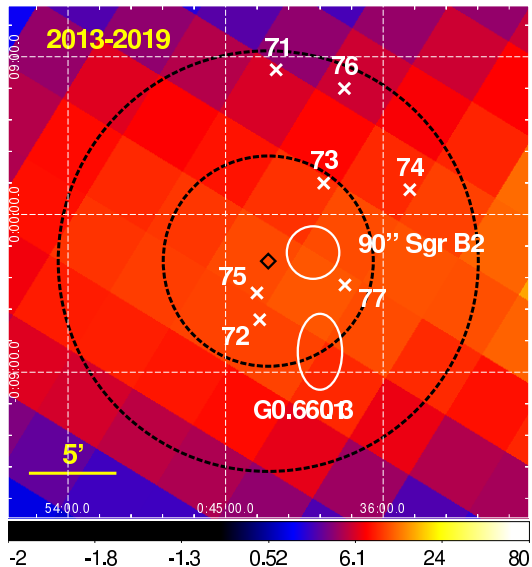}
\caption{The close view of the T2 \sgrb\ image from Fig.~\ref{fig:map_2}. Black dashed circles with radii $R=6-12'$ are centered at the \intl\ \sgrb\ position (diamond) obtained for the T2 data. Solid circle and ellipse are $R=90''$ \sgrb\ region and G0.66--0.13 feature from \citep{Zhang2015}, respectively. Point sources reported in \citep{Hong2016} are shown with crosses.}
\label{fig:point}
\end{figure}

The remaining \sgrb\ emission could also be explained by an integrated flux of the unresolved X-ray sources in this region. To check this, we sum up the fluxes from the known X-ray sources in the \sgrb\ region.

\cite{Zhang2015} presented the results of the \nus\ observations of \sgrb\ with the coverage region almost matching the \intl\ region from which the \igr\ emission was collected (a circle with radius equals \intl\ $FWHM = 12'$). In this region, \nus\ observed only three bright sources in the 10--40~keV energy band: the \sgrb\ core (circle region with $R = 90''$), the extended feature G0.66--0.13, and the point source CXOUGC~J174652.9--282607 \citep[see Fig.~1 in][]{Zhang2015}. We also considered point X-ray sources detected by  \nus\ during the hard X-ray survey of the GC region \citep[see Table~5 in][]{Hong2016} with maximum offset $11\farcm4$ from the T2 \intl\ \sgrb\ position (see Sect.~\ref{sec:spec}). Fig.~\ref{fig:point} demonstrates the \nus\ source positions on the 30--80~keV \intl\ map. We recalculated the fluxes of each \nus\ source in the \ib\ energy band 25--50~keV and summed them to obtain the total flux $\sim3.4\times10^{-12}~\rm erg~cm^{-2}~s^{-1}$ which turned out to be $\sim2$ times lower than the flux $F_{\rm 25-50~keV}=(7.2\pm0.7)\times10^{-12}~\rm erg~cm^{-2}~s^{-1}$ measured by \intl. We conclude that the integrated flux from known X-ray sources does not account for more than half of the observed \sgrb\ emission after 2011.

\section{Summary}
\label{sec:sum}

Thanks to the regular \intl\ observations of the GC region, we have a unique possibility to trace the long-term evolution of the \sgrb\ hard X-ray emission, broadly accepted as the X-ray reflected emission from the past flare activity of \sgra.

We constructed one-year averaged maps of the GC region observed by \intl\ in 2003--2020 that confirm fading of the hard X-ray emission. The sky maps after 2011 demonstrate a significant emission from the \sgrb\ position. We constructed the \sgrb\ light curve for the 2003--2019 time period and found that the light curve is better described with a piece-wise function than a linear one. The fitting procedure with the piece-wise function showed that in $2011\pm3$~yr, there is a transition in the 30--80~keV \sgrb\ flux light curve from fading to constant. The existence of the break in the \sgrb\ time evolution may indicate a change in the emission generation mechanism. However, the possibility that the whole light curve is consistent with the linear decrease associated with the reflection scenario cannot be completely ruled out.

The spectral analysis showed that there is no difference in spectral shape before and after $\sim2011$. Both spectra are in good agreement with a power-law model with a high-energy cutoff at $\sim43$~keV. Also, both spectra support the reflection scenario of an X-ray flare with $\Gamma\sim2.3$. The fading emission observed before 2011 is well explained by the reflection scenario, while the nature of the remaining emission is still unclear.

The low-energy CR protons model with primary power-law slope $s=2.7\pm0.5$ describes the 2011--2019 spectrum well, but the estimated CR ionization rate $\zeta_{\rm H}\sim5.6\times10^{-14}$~H$^{-1}$~s$^{-1}$ is one order of magnitude higher than the GC value $\zeta_{\rm H}\sim(1-3)\times10^{-15}$~H$^{-1}$~s$^{-1}$ \citep{Goto2011}. Therefore this scenario is unfavorable for the emission observed after 2011. Part of this emission could be a result of the composition of unresolved point sources. The known \nus\ sources could explain only about 50\% of the observed \intl\ flux in 25--50~keV and do not dominate in the 2011--2019 \sgrb\ flux. Another significant part of the remaining \sgrb\ emission could be caused by multiple scatterings in the reflection scenario. Further observations are needed to precisely investigate the \sgrb\ region at low and high X-ray energies.

\section*{Acknowledgments}
This work is based on observations with \intl, an ESA project with instruments and the science data centre funded by ESA member states (especially the PI countries: Denmark, France, Germany, Italy, Switzerland, Spain), and Poland, and with the participation of Russia and the USA. We are grateful to Eugene Churazov, Ildar Khabibullin and Vincent Tatischeff for the useful comments and suggestions. E.K. and A.L. acknowledge support from the Russian Foundation for Basic Research (grant 19-32-90283). R.K. acknowledges support from the Russian Science Foundation (grant 19-12-00369). M.C. acknowledges financial support from the Centre National d'Etude Spatiales (CNES).

\section*{Data Availability}
The \intl\ data underlying this article are publicly available at \url{http://www.isdc.unige.ch/}.

\bibliographystyle{mnras}
\bibliography{biblio}

\bsp	
\label{lastpage}
\end{document}